\documentclass[twocolumn, a4]{revtex4}
\usepackage{graphics}
\usepackage{amssymb}

\def\ee{\end{equation}}

\def\be{\begin{equation}}
\def\ee{\end{equation}}
\def\beq{\begin{equation}}
\def\eeq{\end{equation}}
\def\bea{\begin{eqnarray}}
\def\eea{\end{eqnarray}}

\def\beq{\begin{equation}}
\def\eeq{\end{equation}}
\def\bea{\begin{eqnarray}}
\def\eea{\end{eqnarray}}

\usepackage{bm}
\usepackage{graphicx}

\begin{document}
\bibliographystyle{unsrt}
\title[Probing quantum coherence in qubit arrays]
{Probing quantum coherence in qubit arrays}
\author{J. Almeida$^{1,2}$, P. C. de Groot$^{3,4,5}$, S. F. Huelga$^{1,2}$, A. M. Liguori$^{1,2}$ and M. B. Plenio$^{1,2}$}
\address{$^1$ Institut f{\"u}r Theoretische Physik, Universit{\"a}t Ulm, Albert Einstein Allee 11, D-89069, Ulm, Germany}
\address{$^2$ Institute for Integrated Quantum Science and Technology,
Albert-Einstein-Allee 11, University Ulm, D-89069 Ulm, Germany}
\address{$^3$ Kavli Institute of Nanoscience, Delft University of Technology, 2600 GA Delft, The Netherlands}
\address{$^4$ Fakult\"{a}t f{\"u}r Physik, Ludwig-Maximilians-Universit\"{a}t, Schellingstrasse 4, D-80799 M\"{u}nchen, Germany}
\address{$^5$ Max-Planck-Institut f{\"u}r Quantenoptik, Hans-Kopfermann-Strasse 1, D-85748 Garching, Germany}

\begin{abstract}
We discuss how the observation of population localization effects in
periodically driven systems can be used to quantify the presence of
quantum coherence in interacting qubit arrays. Essential for our
proposal is the fact that these localization effects persist beyond
tight-binding Hamiltonian models. This result is of special practical
relevance in those situations where direct system probing using
tomographic schemes becomes infeasible beyond a very small number of
qubits. As a proof of principle, we study analytically a Hamiltonian
system consisting of a chain of superconducting flux qubits under the
effect of a periodic driving. We provide extensive numerical support
of our results in the simple case of a two-qubits chain. For this
system we also study the robustness of the scheme against different
types of noise and disorder. We show that localization effects
underpinned by quantum coherent interactions should be observable
within realistic parameter regimes in chains with a larger number of
qubits.
\end{abstract}

\newpage

\maketitle

\section{Introduction}

Transport processes are of fundamental importance in a wide variety of
physical and biological systems, ranging from the actual motion of
particles on a lattice \cite{dunlap86,holthaus96} to the transfer of
classical and quantum information across spin or harmonic chains
\cite{chain1,chain2,chain3}. Relevant for our purposes, there exist
specific features of the transport process that are intrinsically
linked to the dynamics of the chain and in particular to whether or
not the chain elements can interact coherently \cite{review2}. In the
mid-$80$'s the motion of a charged particle on a one-dimensional
lattice under the influence of a time-dependent electric field was
studied and shown to exhibit {\em dynamic localization} (DL)
\cite{dunlap86}. The canonical situation to illustrate this phenomenon
is provided by an infinite linear chain of sites along which a charged
particle moves under the combined influence of a nearest-neighbour
exchange interaction and a time-dependent external driving. In that
setting, it was found that the mean-square displacement of the
particle as a function of the field modulation $E_1$, rather than
exhibiting a diffusive behaviour, does not grow without bounds but
oscillates sinusoidally. A related phenomenon is the so called {\em
  coherent destruction of tunneling} (CDT), initially formulated in
dissipationless conditions for a symmetric, externally driven,
double-well potential \cite{gross} and subsequently also studied in a
dissipative environment \cite{MilenaPR1998} (and references
therein). Both DL and CDT are genuine manifestations of coherent
quantum effects resulting from the interference between different
transition paths that leads to the selective inhibition of transport
\cite{japan}. In contrast, in the classical case, and from an
initially localized state, an equilibrium state would be attained in
which neighbouring sites would be equally populated. \\ Ample
experimental evidence supports the existence of both types of
localization effects in a variety of systems. DL has been observed in
Rydberg atoms, where the localization regime is characterized by a
"freezing" of the width of the wave packet with respect to the Rydberg
levels \cite{blumel}; in driven quantum wells or semiconductor
superlattices, where a suppression of the conductance was observed
\cite{keay}; in ultracold atoms interacting with a standing wave of
near-resonant light, where this phenomenon was found in the
suppression of momentum \cite{moore-raizen, bharucha-raizen}. There
exist also experimental proposals to use CDT as a means to control the
dynamics of ultracold atoms in optical lattices \cite{CreffieldPRL99}
and to create long-distance entanglement between atoms with the
possibility to use it in the implementation of quantum logical gates
\cite{CreffieldPRL99, GalvePRA79}. CDT has also been recently observed
in both noninteracting \cite{cdt1} and interacting systems
\cite{cdt2}.  Further, motivated by the desire to study the effects of
quantum coherence and dephasing noise and the interplay of the two on
transport processes in biological systems \cite{NJP}, recently the
detection of dynamic localization was proposed as a way to demonstrate
the possible existence of coherence effects in ion channels
\cite{plenio-vaziri2010}, i.e. protein complexes that regulate the
flow of particular ions across the cell membrane and that are
essential for a wide variety of cellular functions.

Here we extend this work and discuss the possibility of observing
localization effects beyond the canonical setting, including
deviations from a strict tight-binding Hamiltonian as well as the
inclusion of non-Hamiltonian (noisy) effects. We will show that
signatures of localization effects can still be observed in this case
and apply these results to the problem of qualitatively witnessing
quantum coherence in an interacting chain of superconducting qubits.

\section{Renormalization of intra-qubit interactions by means of an external modulation}

Motivated by specific qubit realizations in the solid state, we analyze an array of interacting qubits subject to a Hamiltonian of the form ($\hbar=1$):

\begin{equation}
\label{H}
H_0= \sum_{k=1}^N
\frac{\omega_k}{2}\sigma_k^z +\sum_{k=1}^{N-1} J_{k,k+1}\sigma_k^x\sigma_{k+1}^x,
\end{equation}
with $N$ denoting the number of qubits in the chain, $\omega_k$ the {\em site} energies for each qubit, and $J_{k,k+1}$ the coupling between
neighbouring qubits $k$ and $k+1$. In the presence of a time-dependent
external driving of the form

\begin{equation}
\label{eq:extH}
H_{\rm ac}(t)=\frac{1}{2} \sum_{k=1}^N k \cdot E_{\rm ac}\cos(\omega t) \sigma_k^z
\end{equation}
the Hamiltonian $H(t)=H_0+ H_{\rm ac}(t)$ of the chain reads

\begin{equation}
\label{H-fin}
H=\frac{1}{2} \sum_{k=1}^N \big(\omega_k + k E_{\rm ac}\cos(\omega t)\big)\sigma_k^z +
\sum_{k=1}^{N-1} J_{k,k+1}\sigma_k^x\sigma_{k+1}^x.
\end{equation}
With the substitution $\sigma_k^x=\sigma_k^+ + \sigma_k^-$, the
Hamiltonian above can be rewritten as the sum of three contributions,

\begin{equation}
\label{eq:hamPP}
H(t)=H_z(t) + H_1 + H_2
\end{equation}
with,

\begin{eqnarray}
\label{eq:hamZ12}
H_z(t)\equiv
\frac{1}{2} \sum_{k=1}^N \big(\omega_k + k E_{\rm ac}\cos(\omega t)\big)\sigma_k^z\\
H_1\equiv
\sum_{k=1}^{N-1} J_{k, k+1} \big( \sigma^+_k \sigma^-_{k+1} + \textrm{h.c.} \big)\\
H_2\equiv
\sum_{k=1}^{N-1} J_{k, k+1} \big( \sigma^+_k \sigma^+_{k+1} + \textrm{h.c.} \big).
\end{eqnarray}
Defining the total excitation number operator as
\begin{equation}
\hat{N}\equiv \sum_{k=1}^{N} \sigma_k^+\sigma_k^-
\end{equation}
it is easy to see that $[H_z(t), \hat{N}]=[H_1, \hat{N}]=0$ while
$[H_2, \hat{N}]\ne 0$. For this reason, the term $H_1$ is usually
referred to as an exchange interaction, in the sense that it allows
for a hopping of the excitations within the chain, but it does not
create nor annihilate them. This is the canonical interaction in
previous studies of dynamical localization in systems that can be
modelled with a tight-binding Hamiltonian \cite{dunlap86}. \\

In the following lines we will show that the interactions described by
the terms $H_1$ and $H_2$ can be indeed enhanced or inhibited
separately by the proper tuning of frequency $\omega$ and amplitude
$E_{\rm ac}$ of the external field.\\
To gain insight into the problem, it is convenient to move to an
interaction picture with respect the time-dependent term
$H_z(t)$. That is, we first define

\begin{equation}
\label{eq:U0}
U_0(t)\equiv \exp_{+}\left \{-i\int_0^{t} d\tau H_z(\tau)\right \}=
\exp \left \{-i\int_0^{t} d\tau H_z(\tau)\right \}
\end{equation}
where we have made use that $H_z(t)$ commutes with itself at different
times to write the last equality above. Computing explicitly the
integral above and taking into account that the operators acting on
different sites commute, we have that
\begin{equation}
U_0=\prod_{k=1}^{N}
\exp\left \{ -i\frac{ \omega_kt}{2}\sigma_k^z
-i\frac{k E_{\rm ac}}{2\omega}\sin(\omega t)\sigma_k^z
 \right \}
\label{eq:U0}
\end{equation}\\
Hence, the interaction picture
Hamiltonian of our chain can be written as
\begin{equation}
H^{\prime}(t)\equiv U_0(t)^{\dagger}(H_1+H_2)U_0(t)
\equiv H^{\prime}_1(t) + H^{\prime}_2(t),
\end{equation}
where we have defined $H^{\prime}_1(t) \equiv
U_0(t)^{\dagger}H_1U_0(t)$ and $H^{\prime}_2(t) \equiv
U_0(t)^{\dagger}H_2U_0(t)$. Now, making use of eq.  (\ref{eq:U0}) and
the Jacobi-Anger expansion $e^{iz\sin(\phi)} =
\sum_{n=-\infty}^{\infty} J_n(z)e^{in\phi}$, where $J_n(x)$ are the
Bessel functions of first kind, we can proceed further and evaluate
the form of these terms explicitly as:

\begin{widetext}
\begin{eqnarray}
H^{\prime}_1(t)=
\sum_{k=1}^{N-1} J_{k,k+1} \sigma^+_k\sigma^-_k
\exp\left \{i(\omega_k-\omega_{k+1})t -i \frac{E_{\rm ac}}{\omega} \sin(\omega t) \right\} +\textrm{h.c.}\\ \nonumber
=
\sum_{k=1}^{N-1} J_{k,k+1} \sigma^+_k\sigma^-_k e^{i(\omega_k-\omega_{k+1})t}
\sum_{n=-\infty}^{\infty} J_n\left(\frac{E_{\rm ac}}{\omega}\right)
e^{inwt} + \textrm{h.c.}
\end{eqnarray}

\begin{eqnarray}
H^{\prime}_2(t)=
\sum_{k=1}^{N-1} J_{k,k+1} \sigma^+_k\sigma^+_k
\exp\left \{i(\omega_k+\omega_{k+1})t  + i \frac{E_{\rm ac}(2k+1)}{\omega} \sin(\omega t) \right\} +\textrm{h.c.}\\ \nonumber
=
\sum_{k=1}^{N-1} J_{k,k+1} \sigma^+_k\sigma^+_k
e^{i(\omega_k+\omega_{k+1})t}
\sum_{n=-\infty}^{\infty} J_n\left(\frac{E_{\rm ac}(2k+1)}{\omega}\right)
e^{inwt} + \textrm{h.c.}
\end{eqnarray}
\end{widetext}
For the sake of clarity we will consider in the following the case of
a homogeneous chain with $\omega_k=\omega_0$ and $J_{k,k+1}=J$ for all
values of $k$. Under this assumption the Hamiltonian $H^{\prime}(t)$
can be written as:
\begin{equation}
\label{eq:effHam}
H^{\prime}(t)=\sum_{k=1}^{N-1}
g(t) \sigma_k^+\sigma_{k+1}^-
+g^{\prime}_k(t) \sigma_k^+\sigma_{k+1}^+
+\textrm{h. c.}
\end{equation}
with time-dependent renormalized couplings $g(t)$ and
$g^{\prime}(t)$ defined as

\begin{equation}
\label{eq:gprime_a}
g(t)\equiv
J \sum_{n=-\infty}^{\infty} J_n\left(\frac{E_{\rm ac}}{\omega}\right)
e^{inwt}
\end{equation}

\begin{equation}
\label{eq:gprime_b}
g^{\prime}_k(t)\equiv
J \sum_{n=-\infty}^{\infty} J_n\left(\frac{E_{\rm ac}(2k+1)}{\omega}\right)
e^{i(2\omega_0+nw)t}
\end{equation}
In the regime where the tunneling frequency of the qubits is much
smaller than the frequency of the driving field, that is $J\ll
\omega$, we can invoke the rotating wave approximation in the series
above and neglect those terms that rotate faster than $J$. In
particular, for $g(t)$ this means that only the non-rotating term with
$n=0$ survives and we can write

\begin{equation}
g(t)\equiv g=J\cdot J_0 \left(\frac{E_{\rm ac}}{\omega}\right).
\end{equation}
Applying the same reasoning to $g^{\prime}(t)$ it follows that the
only possibility to have surviving terms is that a resonance between
$\omega_0$ and $\omega$ ocurrs such that
$2\omega_0+n^{\prime}\omega=0$ for some integer value $n^{\prime}\in
\mathbb{Z}$. In this case, eq.(\ref{eq:gprime_b}) can be further
simplified, yielding

\begin{equation}
g^{\prime}_k(t)\equiv g_k^{\prime}=J\cdot J_{\vert n^{\prime}\vert }
\left(\frac{\vert n^{\prime} \vert E_{\rm ac}(2k+1)}{2\omega_0}\right).
\end{equation}

We therefore see that the effect of the external modulation can be
interpreted as renormalization of the coupling constants, imprinting a
periodic dependence that will lead to a selective inhibition of
transport.  In the following sections we will provide numerical
evidence of the accuracy of the expressions derived above. This type
of localization effect will be later exploited to detect signatures of
coherent interaction in arrays of coupled qubits.

\section{System description}
We will study the persistence of population localization effects
induced by the renormalization of the hopping coupling, as explained
in the previous section, in the dynamics of superconducting qubit
arrays. Superconducting qubits are effective two-level systems with a
controllable transition frequency between their eigenstates, whose
potential to be manufactured lithographically in a controlled manner
and in a variety of geometries makes them a promising candidate for
the implementation of quantum registers and information processers
\cite{devoret2008}. On the other hand, while the fabrication of
structures involving many qubits is indeed feasible \cite{dwave}, its
effective probing, and in particular the verification that the system
does exhibit quantum coherence, is beyond the realm of current
technology even for moderate system size. The current state of the art
is provided by the tomographic analysis and the entanglement
verification of 3 qubit systems \cite{nature1,nature2,nature3}.
\begin{figure}[h!]
\centering
\includegraphics[width=9cm]{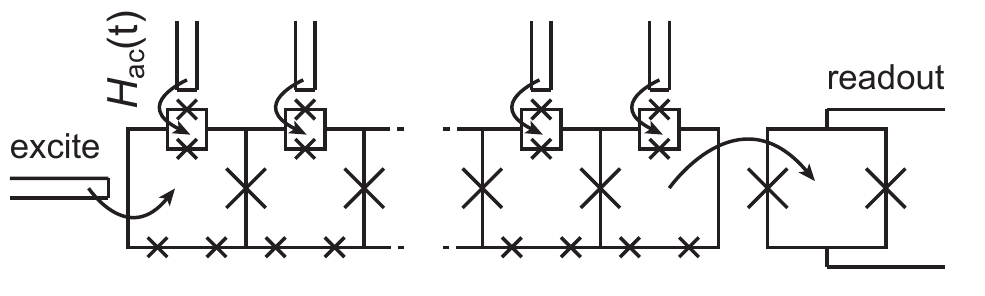}
\caption{Schematic depiction of a superconducting qubit chain with
  controllable injection and readout as well as individual addressing
  for local modulation. By means of population measurements on a
  selected qubit, the presence of quantum coherence is inferred from
  dynamical localization effects, as described in the main text.}
\label{fig:schemepic}
\end{figure}

In general the superconducting qubit Hamiltonian can be written as
$H_{s} = (\epsilon/2)\, \sigma^z + (\Delta/2) \, \sigma^x$. Depending
on the particular qubit realization the parameters $\epsilon$ and
$\Delta$ refer to different variables defining "charge", "phase" or
"flux" qubits. The latter, also called "persistent current qubit",
consists of a superconducting loop interrupted by three Josephson
junctions, two with capacitance $C_1$ and the third with $C_2$
\cite{mooij99,orlando99,devoret2008}. The values of the three
Josephson junctions coupling constants, $E_{J,1}$ corresponding to
capacitance $C_1$ and $E_{J,2}$ to capacitance $C_2$ respectively, are
chosen so that the Josephson part of the Hamiltonian alone defines a
bistable system. At a value of external magnetic flux $\Phi \approx
0.5\Phi_0$ ($\Phi_0=h/2e$ is the superconducting flux quantum), the
system carries either a clockwise or counterclockwise persistent
current, each generating an equal but opposite magnetic flux and
defining the two possible states of this qubit. With an appropriate
choice of the parameters $E_{J,1,2}$ and $C_{1,2}$, the barrier in
phase space separating the left and right current states can be made
low enough so that tunneling between the two classical states can
occur. For the flux qubits $\Delta$ represents the tunneling
amplitude, while the energy bias $ \epsilon=-I_p(\Phi-0.5\Phi_0) $ is
proportional to the detuning $\Phi-0.5\Phi_0$, with $I_p$ the
circulating current. The energies of the ground and first excited
states are thus: $E_{\mp}=\mp \sqrt{\epsilon^2 + \Delta^2}$.

For flux qubits, the most natural implementation for the interaction
is through the mutual inductances between the loops. The flux
generated by one qubit loop, which depends on its internal state, adds
to the total flux picked up by the neighbouring qubits, thereby
changing the energy biases of those qubits. The mutual inductance, and
therefore the strength of the interaction ($J$), depends on the
geometry of the qubit loops, specifically their size and their
proximity to each other. The strength of the coupling can be enhanced
in two steps. First by physically connecting the loops so that their
persistent currents share a common line. In this case the kinetic
inductance of the shared line adds to the geometrical inductance,
where the former term can easily be the dominating part in the total
interaction strength. To reach very strong coupling, for example to
reach the regime $J>\Delta$, a fourth junction can be placed in the
shared line. If the capacitance and Josephson junction coupling
constant of this fourth junction are large compared to parameters of
the qubit junctions, the single-qubit properties are not significantly
altered, while still the interaction strength can be enhanced
dramatically. Coupling strengths of several GHz are easily
achieved. For the sake of the suggested experiments in this paper it
is however not necessary to reach such high coupling values since the
hopping inhibition between the qubits is enhanced in the regime where
the hopping constant is smaller than the driving frequency, i.e. $J\ll
\omega$ (see eqs. (\ref{eq:gprime_a}, \ref{eq:gprime_b}) and
explanations below those expressions).

In the experiment described in reference \cite{paauw-mooij}, it is
demonstrated how the standard flux qubit design, for which only
$\epsilon$ is a tunable parameter, can be extended to also have a
tunable $\Delta$. The main change is that the smallest junction of the
qubit is replaced by a small loop containing one junction in each arm,
i.e. in a SQUID (Superconducting Quantum Interference device)
geometry. The SQUID acts as a single junctions with tunable $E_J$,
tuned by the flux penetrating the small loop. Local control lines can
be used to change the flux through this loop, thereby controlling the
$E_{J,2}$ of the qubit, and thus the
$\Delta$. Fig.~\ref{fig:schemepic} shows a schematic of a possible
implementation of a chain of strongly interacting flux qubits with
tunable tunneling splitting $\Delta$.  When the array of interacting
superconducting flux qubits is operated at their corresponding
degeneracy point the chain is described by a Hamiltonian of the form
eq.(\ref{H}), with $\omega_i=\Delta_i$. Typical values for the
tunneling amplitudes are in the range of 5-20 GHz while nearest
neighbour couplings $J\sim 200$ MHz. The residual next-to-nearest
coupling is smaller than 10 MHz.

As far as the coupling to the environment is concerned, typically, noise sources that are located relatively far away
from the chain couple to multiple or all qubits in the chain, for
example magnetic coils, or some parts of the control and readout
circuits. Noise sources that are located much closer, such as local
control lines for the individual qubits, or microscopic noise sources
in the materials surrounding the qubits, couple only, or mostly, to a
single qubit. In this work we focus on the latter type: we suppose the
array to be in contact with an external environment that acts locally
on each qubit and can lead in principle to both pure dephasing and
dissipation. In our model each qubit is coupled to its local
environment via a spin-boson Hamiltonian of the form

\begin{equation}
H_{SB} = H_{s} + H_{b} +H_{s-b},
\end{equation}
where the bath is modelled as a collection of harmonic oscillators,
$H_b=\sum_k \omega_k \hat{a}_k^\dag \hat{a}_k$ and system and bath
couple linearly through the Hamiltonian \cite{schoen}
\begin{equation}
H_{s-b} = \sigma_z^i \hat{X}_i.
\end{equation}
Here $\hat{X}_i=\sum_k g_k(\hat{a}_k+\hat{a}_k^\dag)$ denotes the
bath's {\em force} operator. In the qubit eigenbasis and using the
same symbols to denote the Pauli matrices in this new frame in order
not to complicate notation, we can derive, using the standard
assumptions, a Markovian master equation $\dot{\rho} = -i[H,\rho]
+ L_{\rm deph}(\varrho) + L_{\rm diss}(\varrho)$ for the qubit array where the
coupling to the local environment is described in terms of Lindblad
terms of the form \cite{tsomokos,RivasPlato,RivasHuelga}
\begin{equation}
\label{Ldeph}
L_{\rm deph}(\varrho)=\gamma_{\rm deph}\sum_{i=1}^N (2\sigma_i^+\sigma_i^-\varrho\sigma_i^+\sigma_i^- -\{\sigma_i^+\sigma_i^-, \varrho\})
\end{equation}
\begin{equation}
L_{\rm diss}(\varrho)=\gamma_{\rm diss}\sum_{i=1}^N (2\sigma_i^-\varrho\sigma_i^+ -\{\sigma_i^+\sigma_i^-, \varrho\}), \label{Ldiss}
\end{equation}
with $\gamma_{\rm deph}$ and $\gamma_{\rm diss}$ denoting the
dephasing and dissipation rate, respectively,
$\sigma_i^\pm=(\sigma_i^x\pm i\sigma_i^y)/2$ acting on the $i$-th
qubit of the chain and $\{\, , \,\}$ the anticommutation
operation. The value of the noise rates in these expressions depends
strongly on the selected qubit operating point via the parameter
$\theta={\rm arctan}(\Delta/\epsilon)$. Measured values for
$T_1=\gamma_{\rm diss}^{-1}$ range from 150 to 500 ns while pure
dephasing times $(T_2^\star)$ are typically around 300 ns. In the
specific case where each qubit in the chain is operated at the
degeneracy point, so that $\epsilon=0$ for every qubit, the pure
dephasing term, which has a rate $\gamma_{\rm deph} \sim \cos(\theta)$
\cite{tsomokos}, cancels out and the chain is subject to dissipative
noise only. This is the first parameter regime that we are going to
analyze in the next section with the aim of unveiling coherent
dynamics through dynamical localization effects in a driven
chain. Later, still operating each qubit at its degeneracy point, we
will relax the constraint of having only dissipation noise and we will
consider possible dephasing effects arising from terms with the form
of eq. (\ref{Ldeph}).

\begin{figure}
\includegraphics[scale=0.6]{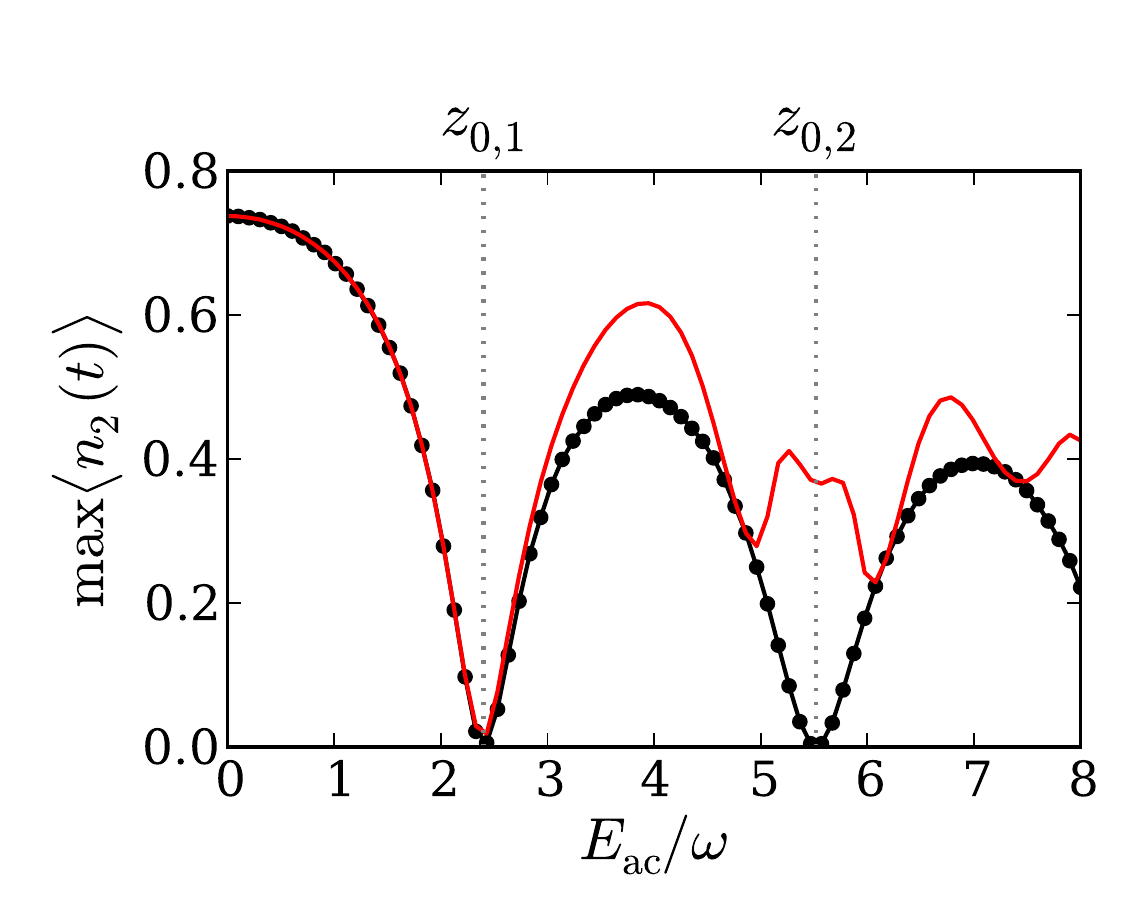}

\caption{ \textbf{Coherent destruction of tunneling}. Simulations corresponding to a $N=2$ chain with $\omega_1=\omega_2=10 \rm GHz$ and
  hopping $J=10 \rm MHz$. The \textit{black dots} have been computed
  with $\omega=0.3 \rm GHz$ and $H(t)=H_z(t)+H_1$ (see definitions in
  the text). The \textit{black line} has been computed with
  $\omega=0.3 \rm GHz$ and $H(t)=H_z(t)+H_1 + H_2$. The \textit{red
    line} has been computed with $\omega=2.0 \rm GHz$ and
  $H(t)=H_z(t)+H_1 + H_2$. The dissipation rate is the same in all
  curves $\gamma_{\rm diss}=1 \rm MHz$. The dashed vertical lines
  represent the first $z_{0,1}\simeq 2.4048$ and second $z_{0,2}\simeq
  5.5201$ zeros of the Bessel function $J_0(z)$.}
\label{fig:dloc}
\end{figure}

\section{Numerical Results}

\begin{figure}
\includegraphics[scale=0.6]{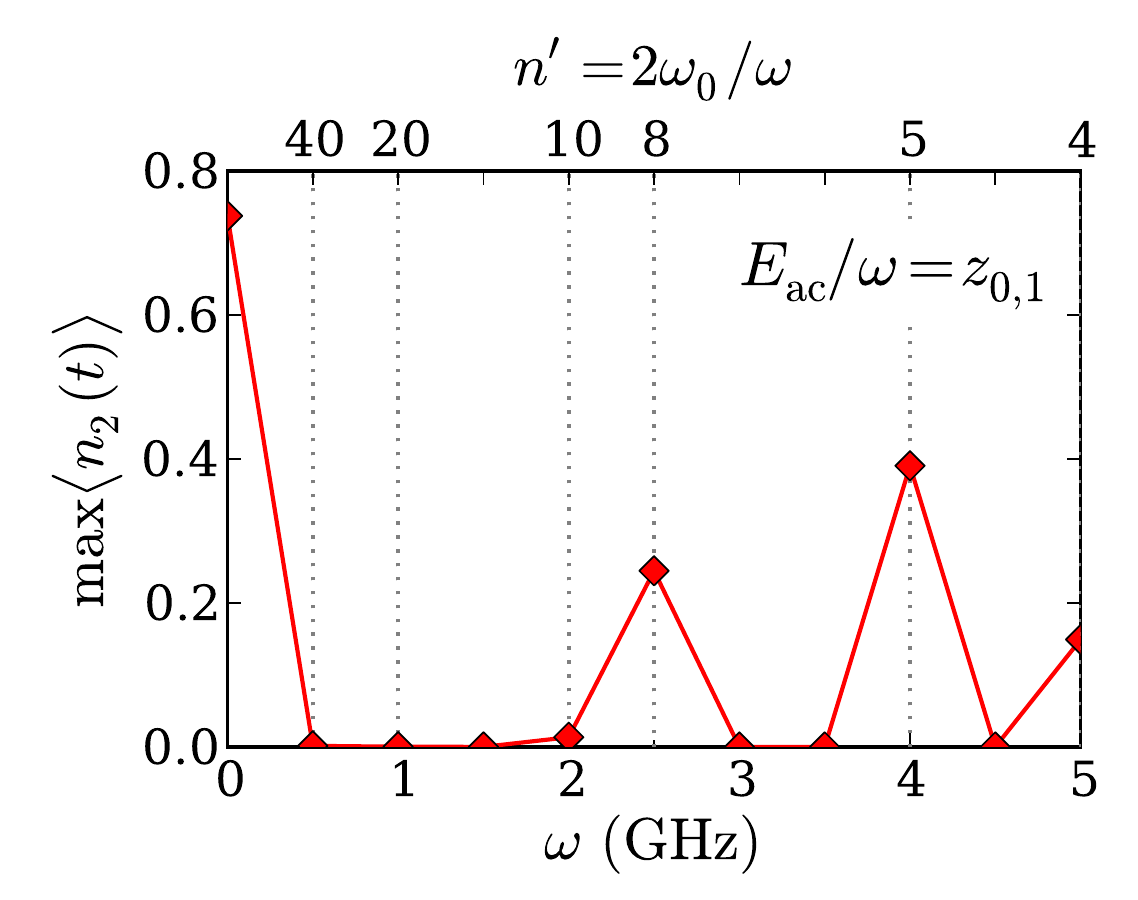}
\caption{\textbf{Resonance conditions}. Simulations for a $N=2$
  chain with $\omega_1=\omega_2=10 \rm GHz$, $J=10\rm MHz$ and a
  dissipation rate $\gamma_{\rm diss}=1\rm MHz$. For each value of
  $\omega$, the electric field is fixed so that $E_{\rm ac}/\omega=z_{0,1}$,
  where $z_{0,1}\simeq 2.4048$ is the first zero of the Bessel function
  $J_0(z)$. The axis on the top represents some integer values
  $n^{\prime}=2\omega_0/\omega$ for which the resonance condition in
  the coupling constant $g^{\prime}$ is fulfilled.}
\label{fig:xx_resonances}
\end{figure}

\begin{figure}
\includegraphics[scale=0.6]{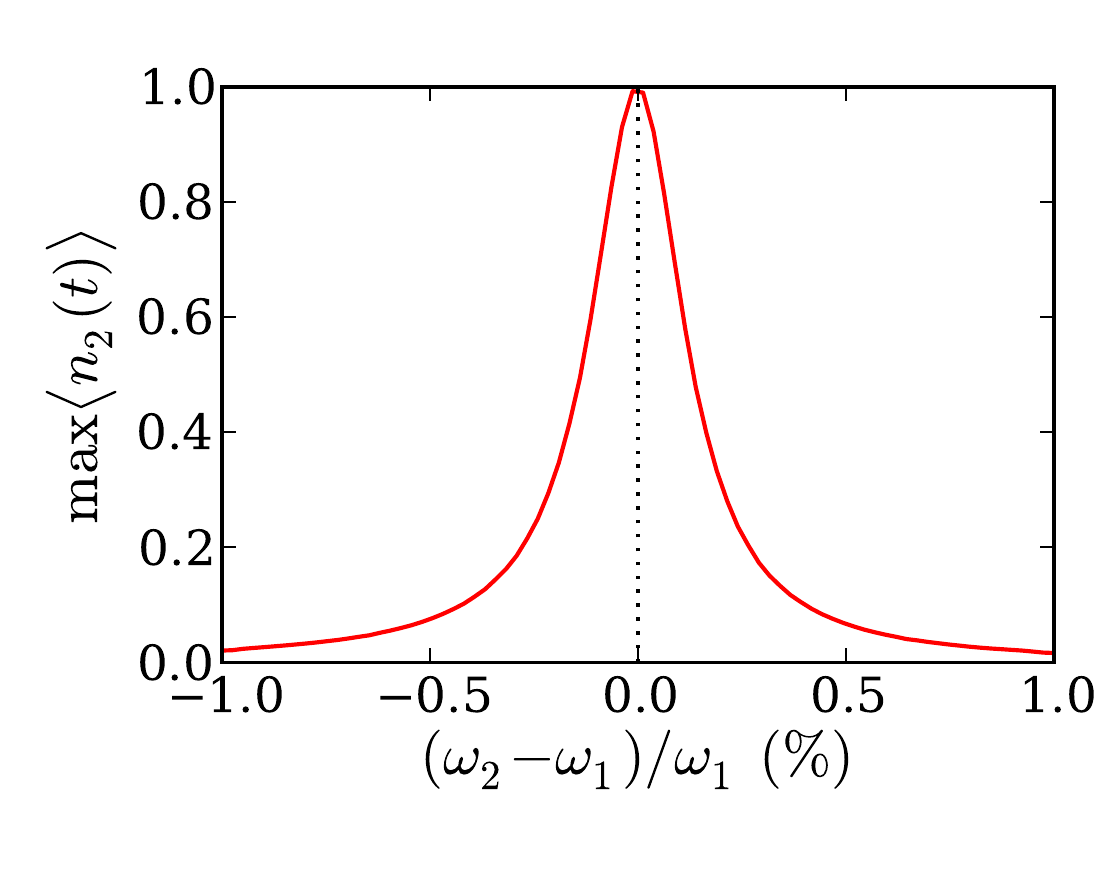}
\caption{\textbf{Effect of frequency inhomogeneities}. Maximum value
  of the population transferred from qubit 1 to qubit 2 in a $N=2$
  chain after a fixed time $t_{\max}=500 \rm ns$. The frequency of the
  external field is $\omega=1.3 \rm GHz$ and its magnitude is fixed
  so that $E_{\rm ac}/\omega=1.2$, far from the closest zero of the
  $J_0(z)$ Bessel function. The parameters of the qubits are
  $\omega_0=10\rm GHz$ and $J=10\rm MHz$. The dephasing and
  dissipation rates have been set to zero for clarity.}
\label{fig:w0_disorder}
\end{figure}

The phenomenon described above concerning the renormalization of the
coupling constants in the interaction Hamiltonian eq.(\ref{eq:hamPP})
under the effect of an external sinusoidal driving field can be
clearly observed in fig.~\ref{fig:dloc}. There we study the dynamics
of a chain consisting in two inductively coupled superconducting
qubits operated at their degeneracy points, so that the (undriven)
system Hamiltonian is given by eq.(\ref{H}). In this case, as
discussed before, the noise is purely dissipative when expressed in
the rotated basis. The system is initialized so that only the first
site is excited. The different curves correspond to the maximum value
of the population that has been transferred from the first to the
second qubit within a sufficiently long time interval ($1\, \mu{\rm
  s}$). The black line corresponds to an off-resonance situation
($\omega_0=10\,{\rm GHz}$, $\omega=0.3 {\rm GHz}$, with no integer
value $n^{\prime}$ such that $2\omega_0+n^{\prime}\omega=0$). In this
situation the coupling of the contribution given by $H_2$ effectively
renormalizes to zero and the dynamics is governed by the tight-binding
term $H_1$. According to expression eq.(\ref{eq:gprime_a}), it is the
Bessel function $J_0(E_{\rm ac}/\omega)$ which governs this behavior
and when its argument $E_{\rm ac}/\omega$ coincides with one of its
zeros, then the hopping between both qubits is suppressed. On the
other hand, the red curve has been computed on a resonance situation
($\omega_0=10\,{\rm GHz}$, $\omega=2.0 {\rm GHz}$, such that
$2\omega_0+n^{\prime}\omega=0$ for $n^{\prime}=-10$). In this
situation both couplings $g$ and $g^{\prime}$ in
eqs.(\ref{eq:gprime_a}, \ref{eq:gprime_b}) are in general different
from zero and the total dynamics is hence more convoluted. Notice that
the fact that both the red and black curves are indistinguishable to
the eye for low values of $E_{\rm ac}/\omega$ is due to the slow
buildup of the Bessel function $J_{n^{\prime}}(z)$ (with
$n^{\prime}=10$) that governs the $H_2$ term. Finally, the black
circles in fig.~\ref{fig:dloc} have been computed using the same
parameters that we used for the off-resonance situation but this time
we explicitly excluded the contribution given by the $H_2$ terms. The
fact that the black line and the black circles superimpose each other
is a clear indication that the term $H_2$ effectively renormalizes to
zero when the field is out of resonance with this term.

\begin{figure}
\includegraphics[scale=0.6]{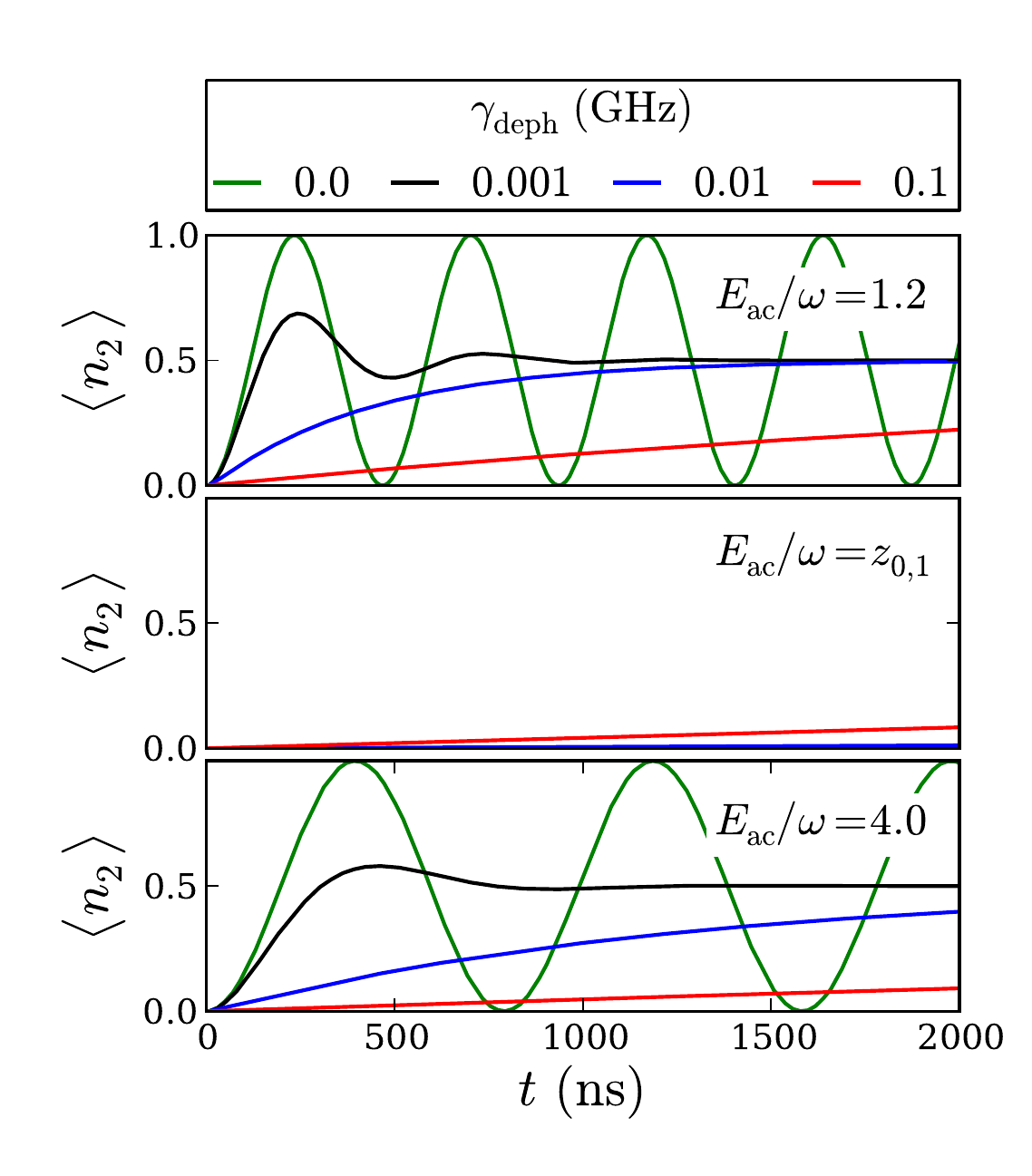}\\
\includegraphics[scale=0.56]{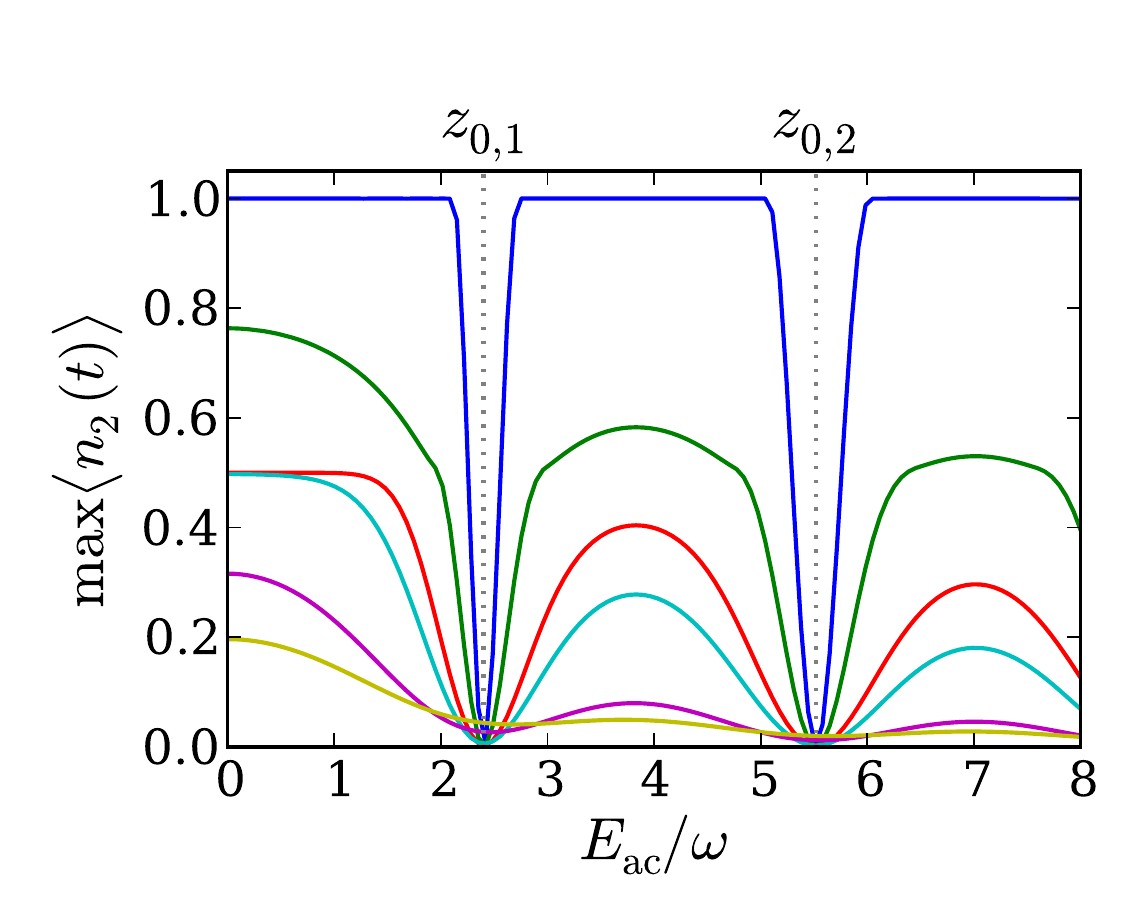}
\caption{\textbf{Effect of pure dephasing noise}.(\textit{top})
  Population transfer from qubit 1 to qubit 2 in a $N=2$ chain and the
  following parameters: $\omega_1=\omega_2= 10\rm GHz$, $J=10\rm MHz$,
  $\omega=1.3\rm GHz$. The dissipation rate has been set to zero
  ($\gamma_{\rm diss}=0.0$) for clarity and a new dephasing rate
  $\gamma_{deph}$ has been included ($\mathcal{L}_{\rm deph}\equiv
  \gamma_{\rm deph}\sum \mathcal{L}_k(\sigma^z)$, with
  $\mathcal{L}_k(\sigma_z)= \sigma_k^z\rho\sigma_k^z -\rho$). Each
  panel corresponds to a fixed value $z=E_{ac}/\omega$ with
  $z_{0,1}\simeq 2.4048$ the first zero of the Bessel function
  $J_0(z)$. (\textit{bottom}) Maximum value of the population
  transferred to the second qubit in the time interval and with the
  same parameters as the figure on \textit{top} and the following
  dephasing rates (in ${\rm GHz}$): $\gamma_{\rm deph}=0.0$
  (\textit{blue}),$\gamma_{\rm deph}=0.001$
  (\textit{green}),$\gamma_{\rm deph}=0.005$
  (\textit{red}),$\gamma_{\rm deph}=0.01$
  (\textit{turquoise}),$\gamma_{\rm deph}=0.05$
  (\textit{cyan}),$\gamma_{\rm deph}=0.1$ (\textit{yellow}) }
\label{fig:gdeph_effect}
\end{figure}

The existence of resonance conditions for the terms contributed by
$H_2$ in expression eq.(\ref{eq:hamPP}) is clearly illustrated in
fig.~\ref{fig:xx_resonances}, which has been evaluated with the same
parameters as in fig.~\ref{fig:dloc}. We have however fixed the ratio
$E_{\rm ac}/\omega=z_{0,1}$ with $z_{0,1}\simeq 2.4048$ the first zero
of the Bessel function of order zero. With this constraint the
population transfer induced by the terms contributed by $H_1$ in
eq.(\ref{eq:hamPP}) is completely suppressed. We have however the
freedom to use different values of the driving field $\omega$ such
that the term $H_2$ is either on- or off-resonance with the frequency
$\omega_0$. We can clearly see in fig.~\ref{fig:xx_resonances} those
values (compatible with the discrete grid used to sample the frequency
$\omega$) where the resonance condition is fulfilled and a peak in the
population transfer appears induced entirely by the terms in $H_2$.\\

As a result, the presence of a {\em correction} to the canonical
Hamiltonian $H_1$ does not hinder the manifestation of localization
effects. By appropriate tuning of the external driving we can select
resonance conditions that lead to inhibition of transport and provide
a fingerprint of the underlying coherent evolution. In
fig.~\ref{fig:w0_disorder} we illustrate the effect of an
inhomogeneous distribution of the tunneling amplitudes of the qubits
within the chain. Not surprisingly, the presence of this sort of
disorder in the array leads to a quick loss of contrast. However, the
fact that the tunneling amplitude of superconducting flux qubits is
actually tunable can provide mechanism to overcome this difficulty by
minimizing or even suppressing the local disorder in the chain
\cite{tunable}. On the other hand, real implementations of interacting
superconducting flux qubits typically result in interaction strengths
differences of the order of few percents between different nearest
neighbors. It is important to stress at this point that this
differences have little effect in our previous discussion since they
do not affect the resonance conditions in eqs.(\ref{eq:gprime_a},
\ref{eq:gprime_b}), that are the key to the hopping inhibition effect
presented in this work.\\
\begin{figure}
\includegraphics[scale=0.6]{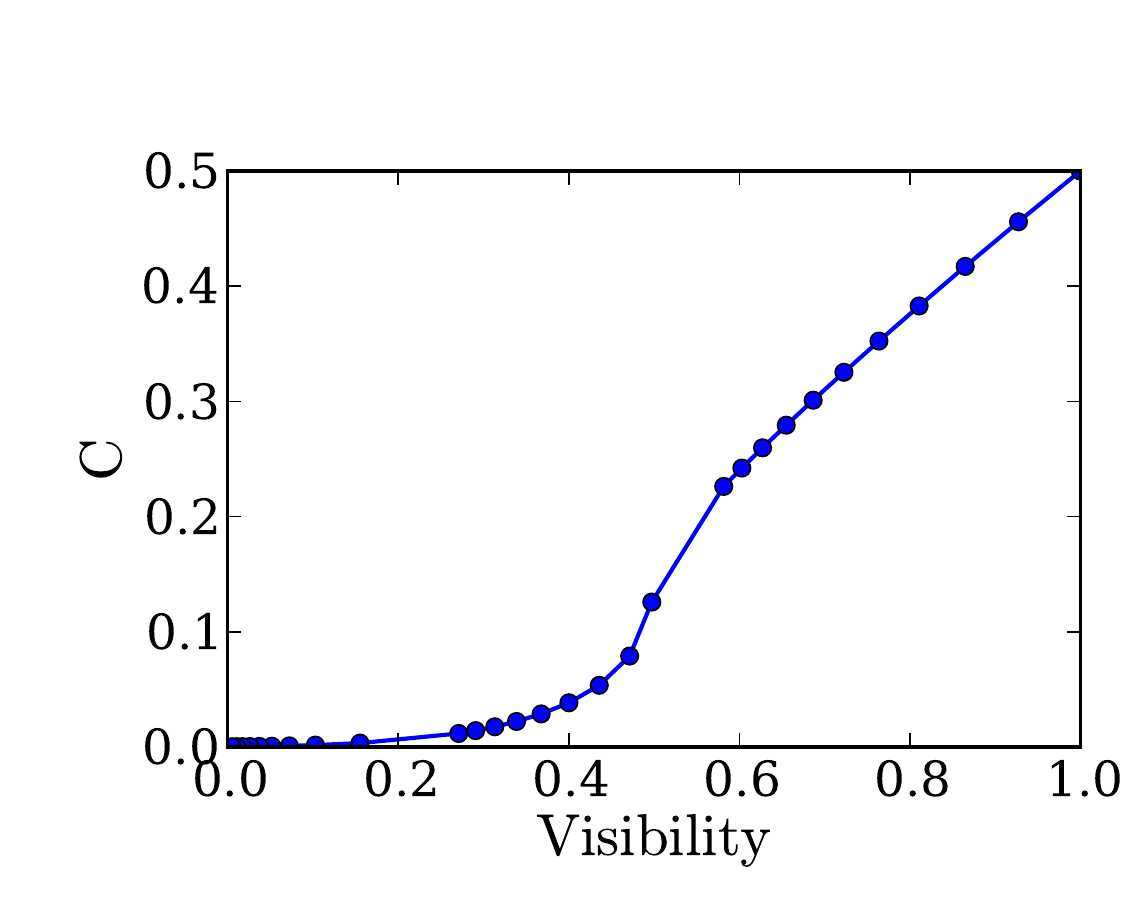}
\caption{\textbf{Quantum coherence in a N=2 chain}. In this graph we
  plot the magnitude $C \equiv \max \sum_{i, j\ne i} {\rm
    abs}\big( \rho_{i,j}(t)\big)$ as a function of the visibility of
  the transport pattern ({\em population fringes} in Fig. 5).  We have
  considered 27 values of the pure dephasing rate in a logarithmic
  scale from $\gamma_{\rm deph}=0.1 {\rm GHz}$ $(C_{\rm max}\simeq 0)$ to
  $\gamma_{\rm deph}=0.0 {\rm GHz}$ $(C_{\rm max}=0.5)$. For
  convenience we have fixed $\gamma_{\rm diss}=0$.}
\label{fig:newcoherences}
\end{figure}

\begin{figure}
\includegraphics[scale=0.6]{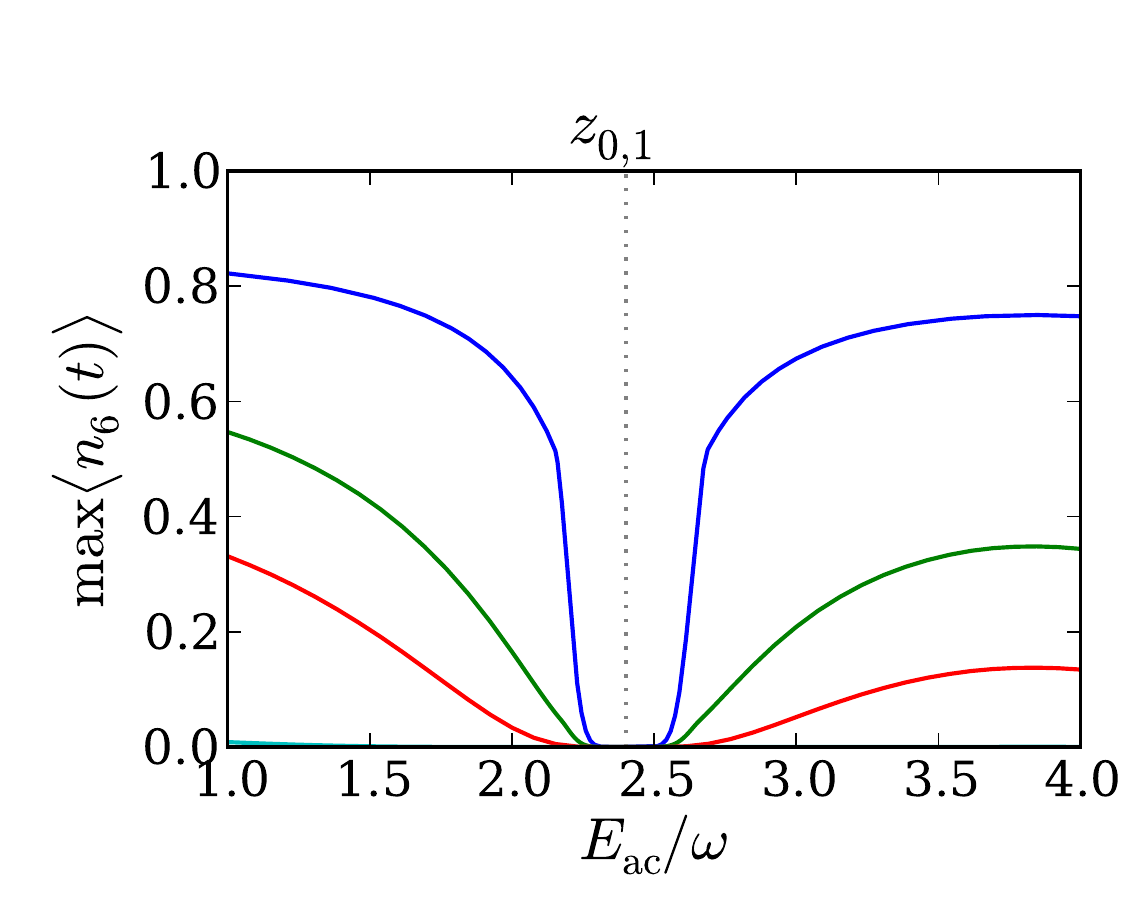}
\caption{\textbf{Hopping inhibition in a N=6 chain}. Maximum value
  of the population transferred from the first qubit to the last one
  in a $N=6$ chain within a time interval of $2.8$ $\mu \rm s $. The
  frequency of the external field is $\omega=1.3 \rm GHz$ and its
  magnitude is fixed so that $E_{\rm ac}/\omega=1.2$, far from the
  closest zero of the $J_0(z)$ Bessel function. The parameters of the
  qubits are $\omega_1=10\rm GHz$ and $J=10\rm MHz$.For this
  computation we have considered dissipation noise only, with the
  following rates (in ${\rm GHz}$): $\gamma_{\rm diss}=0.0001$ (\textit{blue}),
  $\gamma_{\rm diss}=0.0005$ (\textit{green}), $\gamma_{\rm
    diss}=0.001$ (\textit{red}), $\gamma_{\rm diss}=0.005$
  (\textit{turquoise}, slightly visible).}
\label{fig:sixqubits}
\end{figure}
In real implementations of superconducting flux qubits we should also
expect deviations from a purely dissipative model of noise. This
motivates the study of the effect of pure dephasing processes in our
system. To this end we will introduce in the master equation a
Lindblad term of the form given in eq.(\ref{Ldiss}), with dephasing
rate $\gamma_{\rm deph}$. To single out the effect of pure dephasing,
we will take a vanishingly small dissipation rate, that is
$\gamma_{\rm diss}=0$.\\ In fig.~\ref{fig:gdeph_effect} we have
studied how the transference of population in a chain is affected by
pure dephasing noise terms. In
fig.~\ref{fig:gdeph_effect}(\textit{up}) we plot the dynamics of the
population for different values of the dephasing rate, $\gamma_{\rm
  deph}=\{0.0, 0.001, 0.01, 0.1\}{\rm GHz}$. It is worth remarking
that the inclusion of these terms in the master equation leads to an
asymptotic stationary state that is maximally mixed
\cite{ejp}. However, the transient dynamics still provides valuable
information. It is remarkable to notice that even though the dephasing
rapidly destroys the coherent oscillations of the population, still
the renormalization of the hopping rate survives for surprisingly
large values of $\gamma_{\rm deph}$. This fact can be clearly
appreciated in the panel with $E_{\rm ac}/\omega=z_{0,1}$ where the
hopping is strongly inhibited and still noticeable for dephasing rates
as high as $\gamma_{\rm deph}=100$ ${\rm MHz}$. In
fig.~\ref{fig:gdeph_effect}(\textit{down}) we have plotted the same
information in a more compact format. In this graph the effect of
dephasing can be clearly seen to reduce the visibility of the
population transfer oscillations (as a function of $E_{\rm ac}/\omega$
). Nonetheless, as stated above, the hopping modulation is quite
visible also in this graph even for large values of the dephasing
rate.\\ 

The qualitative relation between the properties of the transport
dynamics and the coherence in the system is illustrated in
fig.~\ref{fig:newcoherences}. In this figure we have represented the
coherence of the chain, quantified by the sum of the off diagonal
elements of the density matrix $C\equiv\max \sum_{i, j\ne i} {\rm
  abs}\big( \rho_{i,j}(t)\big)$, as a function of the visibility of
the {\em population fringes} depicted in fig.~\ref{fig:gdeph_effect}.
For the sake of clarity we have used a purely dephasing noise since it
uniquely affects the coherence terms of the density matrix. The main
conclusions are however unaffected by including dissipation terms. We
can see in this graph that the degree of quantum coherence of the
system can be properly quantified by the transport schemes proposed in
this work. For small values of the dephasing rate, the relation
between coherence and visibility is indeed essentially linear. As a
result, population measurements alone would allow, in the presence of
a tunable driving, to detect signatures of quantum coherence in the
system.  The scalability of the procedure is illustrated in
fig.~\ref{fig:sixqubits} for an array of $N=6$ qubits, a system size
that is currently untractable with tomographic schemes. As opposed to
the previous graphs computed for $N=2$, now the one-excitation sector
contains more than two eigenstates and hence the population dynamics
is affected by more than one characteristic frequency. This results in
a more convoluted dynamics and different possible protocols to measure
the hopping inhibition. For this graph we have chosen to start with a
chain where only the first qubit is initially on its excited
configuration, we measure then the maximum population transferred to
the last qubit within a time interval of $2.8 \mu \textrm{s}$, long
enough to allow the initial excitation to reach the end of the
chain. Notice in this graph that the perfect hopping inhibition is
again only achieved at the zeros of the corresponding Bessel
functions. The fact that the hopping seems to vanish in an extended
interval around this point is only a consequence of the interplay
between the finite time interval used to perform the population
measure and the arrival time required for a wavepacket created at the
begining of the chain to reach its end (see ref.\cite{GalvePRA79} for
a thorough study of this topic), which increases as the hopping rate
decreases.

\section{Conclusions}
To summarize, we have analyzed the persistence of localization effects beyond exact tight binding Hamiltonians and beyond a closed system description.
Introducing an a.c. interaction term of the form
of eq.(\ref{eq:extH}), we have seen that the original ZZ coupling, which provides the natural model for the actual coupling in superconducting architectures, can be effectively mapped
into the Hamiltonian eq.(\ref{eq:effHam}) where the excitation-preserving and
non-preserving terms are affected by two different renormalized
couplings $g$ and $g^{\prime}$. We have seen that these effective
couplings are determined by certain resonance conditions (for the
coupling $g$ there is always a resonance for $n^{\prime}=0$, for
$g^{\prime}$ the stronger condition $2\omega_0+n^{\prime}\omega=0$ is
required) and, given that the resonance conditions are fulfilled, by
the arguments of the Bessel functions that define these couplings. In
conclusion, we have proposed a method that allows to tune
independently two kinds of interaction of very different
nature but with the common feature of leading to localization phenomena. These are shown to be useful for witnessing
the coherent behaviour of coupled qubit arrays. As a proof of principle, using typical parameter regimes in chains of superconducting flux qubits,
we have shown that transport inhibition can be qualitatively linked to the degree of coherence in the system.
This type of experiments, which involve population measurements only, can provide a benchmark for quantum behaviour in systems whose complexity makes them unsuited for detailed
tomographic analysis, ranging from arrays of self-assembled quantum dots \cite{dots} to coupled nanomagnets \cite{magnets}.\\

\section{Acknowledgements}
Financial support from the EU Integrated Project Q-Essence, STREP actions CORNER and PICC and the Humboldt Foundation is gratefully acknowledged.
We thank H. Mooij, A. Rivas and B. R\"othlisberger for helpful discussions and their comments on the precursor of this manuscript and to A. Berm\'udez and M. Paternostro
for their feedback on the current manuscript.

\end{document}